\documentclass[rmp,superscriptaddress,10pt,aps]{revtex4-2}

\usepackage{titlesec}
\usepackage{amssymb}
\usepackage{amsmath}
\usepackage{graphicx}
\usepackage[dvipsnames]{xcolor}
\usepackage{bm}
\usepackage{amsthm}
\usepackage{dsfont}
\usepackage{fancyhdr}
\usepackage{hyperref}
\hypersetup{
    colorlinks=true,
    linkcolor=blue,
    filecolor=blue,
    urlcolor=blue,
    citecolor=green
}

\usepackage[a4paper]{geometry}
\setcitestyle{super}

\titlespacing*{\section}{0.ex}{4ex}{0.ex}

\makeatletter
\def\frontmatter@above@affiliation@script{\addvspace{8.0\p@}}
\makeatother

\DeclareMathOperator{\dist}{dist}
\newcommand{\avg}[1]{\big\langle #1 \big\rangle}

\begin{document}

\title{Indirect Influence on Network Diffusion}

\author{\large Llu\'is Torres-Hugas}
        \affiliation{Departament d'Enginyeria Inform\`{a}tica i Matem\`{a}tiques, Universitat Rovira i Virgili, Tarragona, Catalonia, Spain}
\author{\large Jordi Duch}
        \affiliation{Departament d'Enginyeria Inform\`{a}tica i Matem\`{a}tiques, Universitat Rovira i Virgili, Tarragona, Catalonia, Spain}
\author{\large Sergio G\'{o}mez}
        \email{sergio.gomez@urv.cat}
        \affiliation{Departament d'Enginyeria Inform\`{a}tica i Matem\`{a}tiques, Universitat Rovira i Virgili, Tarragona, Catalonia, Spain}

\maketitle

\vspace{-5ex}

\section*{\large{Abstract}}

{\fontfamily{cmss}\selectfont
Models of network diffusion typically rely on the Laplacian matrix, capturing interactions via direct connections. Beyond direct interactions, information in many systems can also flow via indirect pathways, where influence typically diminishes over distance. In this work, we analyze diffusion dynamics incorporating such indirect connections using the $d$-path Laplacian framework. We introduce a parameter, the indirect influence, based on the change in the second smallest eigenvalue of the generalized path Laplacian, to quantify the impact of these pathways on diffusion timescales relative to direct-only models. Using perturbation theory and mean-field approximations, we derive analytical expressions for the indirect influence in terms of structural properties of random networks. Theoretical predictions align well with numerical simulations, providing a phase diagram for when indirect influence becomes significant. We also identify a structural phase transition governed by the emergence of $d$-paths and derive the critical connection probability above which they dramatically alter diffusion. This study provides a quantitative understanding of how indirect pathways shape network dynamics and reveals their collective structural onset.

}

\section*{\large{Introduction}}
\label{sec:Introduction}

Diffusion processes on networks are fundamental in understanding a wide range of phenomena in complex networks \cite{masuda2017random}; these include information spreading in social systems \cite{liu2014information}, disease propagation \cite{newman2002spread}, and transport dynamics in physical and biological systems \cite{de2014navigability, gentili2022biological}. These processes are often modeled using the Laplacian matrix of the network \cite{merris1994laplacian}, which captures the dynamics of local interactions based on the structure of the direct connections between nodes.

In some systems, considering only the direct connections within a network fails to capture all diffusion pathways, and underlying mechanisms may alter how the spreading occurs. To address the limitations of simplifying complex networks as basic graphs, various extensions within the framework of higher-order networks have been proposed. These include multiplex networks, multilayer networks, hypergraphs, simplicial complexes, and others \cite{bick2023higher, de2013mathematical, iacopini2019simplicial}.
These efforts to go beyond pairwise interactions often focus on incorporating group interactions. However, they still overlook an important feature of network dynamics where the strength of interaction gradually decreases as the separation between nodes increases in the underlying graph structure.

In particular, this drawback has been emphasized in the diffusion of information in social systems \cite{granovetter1973strength}, where the decaying influence between nodes is necessary to capture the underlying dynamics, making the introduction of indirect connections, or ``weak ties'', particularly relevant. An indirect connection allows a weaker spread of information between two nodes, $i$ and $k$, that are not directly connected but are connected through a third node $j$, in a path of length two. While $j$ is implicitly involved in the process, it does so passively, without being influenced by the transmitted information.
The importance of indirect connections has also been pointed out in other contexts. For example, second-order neighbors in online networks can significantly influence political mobilization and social influence \cite{bond201261}, and in ecological systems, indirect connections among species can shape coevolutionary dynamics and network structure \cite{guimaraes2017indirect}.

One proposed approach to incorporate these indirect connections is through the use of the $d$-path Laplacian and the $d$-path adjacency matrices \cite{estrada2012path}, a natural generalization of the Laplacian and the adjacency matrices
to include higher-order paths and establish the theoretical possibility of non-local diffusion, even demonstrating super-diffusive behavior under certain conditions for single-particle dynamics\cite{estrada2017path, estrada2018path}. Other options to incorporate indirect influence in networks include the multi-hop random walks model\cite{estrada2018random}, and the fractional Laplacian\cite{riascos2014fractional, estrada2021path}, which show important qualitative and quantitative differences with respect to the $d$-path Laplacian approach. Beyond diffusion, indirect influence can also affect other dynamics, such as synchronization\cite{estrada2018long} and consensus models\cite{gambuzza2019second}.

Indirect diffusion has proven effective in modeling real-world scenarios, e.g., capturing the long-range dispersal of pathogens via vectors in epidemics \cite{arias2018epidemics}, showing how peer pressure can shape consensus and leadership in social groups\cite{estrada2013peer}, and accounting for the accelerated diffusion of innovations through socially-close peers in experiments \cite{miranda2024indirect}.
Despite its practical importance, the interplay between network structure and diffusion dynamics remains poorly understood, although it has proven central in other frameworks \cite{gomez2013diffusion,sole2013spectral}.

In this work, we extend standard network diffusion models to quantitatively assess the impact of influence propagating indirectly along network paths. Employing the $d$-path Laplacian framework, we introduce the indirect influence parameter $\zeta$ to precisely measure the change in diffusion timescales caused by these non-local interactions, linking it directly to the spectral properties of the generalized Laplacian. We provide a theoretical foundation for this parameter, deriving analytical expressions for $\zeta$ in random networks using perturbation theory and mean-field approximations, leading to a predictive phase diagram validated by numerical simulations. Furthermore, we uncover and characterize a structural phase transition associated with the emergence of $d$-paths, deriving its critical threshold $p_c(N)$ and revealing its fundamental role in altering diffusion dynamics. This framework offers a robust method for analyzing the contribution of indirect pathways in complex network processes.

\section*{\large{Results}}

\noindent \textbf{Indirect Influence}

\noindent To analyze diffusion dynamics incorporating non-local interactions, we employ the $d$-path Laplacian framework \cite{estrada2012path}. This approach extends the standard Laplacian model by considering pathways of length $d > 1$ between nodes. The framework uses $d$-path adjacency matrices, $A^{(d)}$, where $\left( A^{(d)} \right)_{ij} = a^{(d)}_{ij} = 1$ if the shortest path distance between nodes $i$ and $j$ is exactly $d$, and $0$ otherwise. From these, the corresponding $d$-path Laplacian, $L^{(d)} = K^{(d)} - A^{(d)}$, is constructed, where $K^{(d)}$ is the diagonal matrix of $d$-path degrees $k^{(d)}_i = \sum_{j} a^{(d)}_{ij}$ (see Methods).

\begin{figure}[tb!]
\centering
\setlength{\tabcolsep}{10pt}
\begin{tabular}[t]{cccc}
    \textit{Direct connections} & \multicolumn{2}{c}{\textit{Indirect connections}} & \textit{All connections} \\
    1-path network & 2-path network & 3-path network & path network \\ \\
    \includegraphics[width=0.2\textwidth]{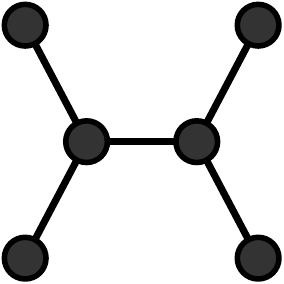} &
    \includegraphics[width=0.2\textwidth]{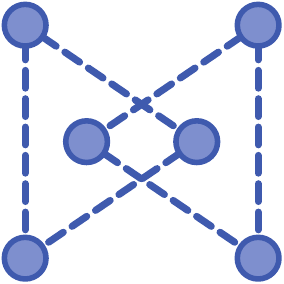} &
    \includegraphics[width=0.2\textwidth]{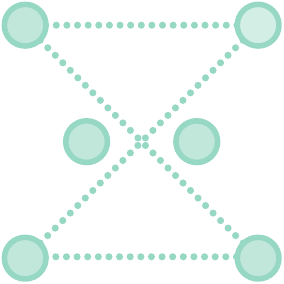} &
    \includegraphics[width=0.2\textwidth]{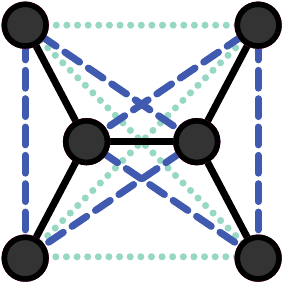}
\end{tabular}
\caption{\fontfamily{cmss}\selectfont{\textbf{Diagrams of $\bm{d}$-path networks and the path network}. The figure shows the $d$-path networks derived from a base network and the corresponding path network that combines all $d$-paths.}}
\label{fig:fig1}
\end{figure}

To capture diffusion across paths of all lengths up to the network diameter $D$, while allowing for influence to decay with distance, we consider the path Laplacian as:
\begin{equation} \label{eq:path_laplacian}
\hat{L} = \sum_{d=1}^{D} d^{-\alpha} L^{(d)},
\end{equation}
where $\alpha \geq 0$ is a parameter controlling the decay rate of influence with path length $d$. Setting $\alpha = 0$ implies uniform influence regardless of distance, while large $\alpha$ recovers the standard local diffusion dynamics dominated by $L^{(1)} = L$. The path Laplacian $\hat{L}$ retains the essential properties of the standard Laplacian, being symmetric and positive semi-definite. Notice that this path Laplacian can be interpreted as the Laplacian matrix of the path network, see Figure~\ref{fig:fig1}, where each link is weighted according to its diffusion rate $d^{-\alpha}$.

%%%%%%%%%%%%%%%%%%%%%%%%%%%%%%%%%%%%%%%%%%%%%%%%%%%%%%%%%%%%%%%%%%%%
% FIGURE 1
%%%%%%%%%%%%%%%%%%%%%%%%%%%%%%%%%%%%%%%%%%%%%%%%%%%%%%%%%%%%%%%%%%%%

The diffusion dynamics on a network governed by the path Laplacian are described by: \begin{equation} \label{eq:diffusion_eqn}
\dot{\mathbf{x}}(t) = -\hat{L}\mathbf{x}(t),
\end{equation}
where $\mathbf{x}(t)$ is the vector of node states at time $t$. Assuming the network is connected, the general solution is given by:
\begin{equation} \label{eq:diffusion_soln}
\mathbf{x}(t) = \sum_{i=1}^N C_i e^{-\hat{\lambda}_i t} \hat{\mathbf{v}}_i
\end{equation}
where $0 = \hat{\lambda}_1 < \hat{\lambda}_2 \leq \cdots \leq \hat{\lambda}_N$ are the eigenvalues and $\hat{\mathbf{v}}_i$ are the corresponding eigenvectors of $\hat{L}$, and $C_i$ are constants determined by initial conditions. The convergence to the equilibrium state, where $x_i = x_j$ for all $i$, $j$, is governed by the second smallest eigenvalue, $\hat{\lambda}_2$, often called the algebraic connectivity or Fiedler value \cite{fiedler1973algebraic, mohar1992laplace}. The characteristic diffusion time is $\tau = 1 / \hat{\lambda}_2$; a larger $\hat{\lambda}_2$ means faster convergence.

In contrast, standard diffusion models only consider direct connections, described by $\dot{\mathbf{x}} = - L \mathbf{x}$, where $L = L^{(1)}$, with a diffusion time $\tau = 1 / \lambda_2$, where $\lambda_2$ is the second smallest eigenvalue of the standard Laplacian $L$. To quantify the impact of including indirect pathways on the diffusion timescale, we introduce the \textit{indirect influence}, defined as the normalized change in the algebraic connectivity:
\begin{equation} \label{eq:zeta_definition}
\zeta = \frac{\hat{\lambda}_2 - \lambda_2}{N}.
\end{equation}
This parameter ranges from 0 to 1, since $N$ is the maximum value that any eigenvalue can take in a Laplacian matrix \cite{fiedler1973algebraic}, and $\hat{\lambda}_2 \geqslant \lambda_2$ since the original 1-path network is a (weighted) spanning graph of the path network \cite{mohar1991laplacian}. A value of $\zeta = 0$ indicates that indirect connections do not alter the diffusion timescale compared to the direct-only model, while $\zeta > 0$ means that indirect pathways accelerate the diffusion process, with larger values indicating a greater impact. This parameter provides a quantitative measure to address when and how significantly indirect connections reshape network diffusion dynamics. \\

\noindent \textbf{Theoretical Approximation of Indirect Influence} \\
\noindent To understand how network structure influences the indirect influence, we derive an analytical expression for it, particularly focusing on general random networks characterized by their size $N$ and expected connection probability $p \equiv \avg{a_{ij}}$. We begin by considering the regime where indirect connections act as a perturbation to the direct diffusion process, which corresponds to a strong decay of influence with distance, $\alpha \gg 1$.

In the large $\alpha$ limit, the path Laplacian $\hat{L}$ is dominated by the $d=1$ term. We start by including the most significant indirect contribution, the $d=2$ term, scaled by a small parameter $\epsilon = 2^{-\alpha} \ll 1$:
\begin{equation} \label{eq:perturbed_L}
\hat{L} = L + 2^{-\alpha} L^{(2)} = L + \epsilon L^{(2)}.
\end{equation}
We can now apply standard perturbation theory to find the change in the second smallest eigenvalue $\lambda_2$ of the standard Laplacian $L$. Let $\lambda_2$ and $\mathbf{v}_2$ be the eigenvalue and corresponding normalized eigenvector of $L$ ($L \mathbf{v}_2 = \lambda_2 \mathbf{v}_2$, $\mathbf{v}_2^T \mathbf{v}_2 = 1$). The perturbed eigenvalue $\hat{\lambda}_2$ of $\hat{L}$ is, to first order in $\epsilon$ (see Methods):
\begin{equation} \label{eq:perturbed_lambda2}
\hat{\lambda}_2 = \lambda_2 + \epsilon \mathbf{v}_2^T L^{(2)} \mathbf{v}_2 + O(\epsilon^2).
\end{equation}
Substituting this into the definition of the indirect influence, we obtain:
\begin{equation} \label{eq:zeta_perturbed}
\zeta \approx \frac{2^{-\alpha} \mathbf{v}_2^T L^{(2)} \mathbf{v}_2}{N},
\end{equation}
to first order in $\epsilon$.
This expression quantifies the indirect influence arising from $2$-paths in the perturbative regime, but it still depends on the specific eigenvector $\mathbf{v}_2$ and the $2$-path Laplacian $L^{(2)}$ of a given network.

To obtain an expression dependent only on average network properties, we employ a mean-field approach to approximate the expectation value of the quadratic form $\mathbf{v}_2^T L^{(2)} \mathbf{v}_2$ over an ensemble of general random networks. Assuming the entries of the eigenvector $\mathbf{v}_2$ are uncorrelated with the entries of $L^{(2)}$ and are independent identically distributed random variables satisfying $\sum_{i=1}^N (\mathbf{v}_2)_i^2 = 1$ and $\sum_{i=1}^N (\mathbf{v}_2)_i = 0$ \cite{fiedler1973algebraic}, we can approximate the expectation:
\begin{equation} \label{eq:quadratic_form_mf_step1}
\begin{split}
\avg{\mathbf{v}_2^T L^{(2)} \mathbf{v}_2} & = \avg{\sum_{i=1}^N k^{(2)}_i {(\mathbf{v}_2)}^2_i} - \avg{\sum_{i=1}^N \sum_{j=1}^N a^{(2)}_{ij} {(\mathbf{v}_2)}_i {(\mathbf{v}_2)}_j} \\ & \approx  N \avg{k^{(2)}_i} \avg{(\mathbf{v}_2)^2_i} - N (N-1) \avg{a^{(2)}_{ij}} \avg{(\mathbf{v}_2)_i (\mathbf{v}_2)_j} .
\end{split}
\end{equation}
Using the properties $\avg{(\mathbf{v}_2)_i^2} = \frac{1}{N}$ and $\avg{(\mathbf{v}_2)_i(\mathbf{v}_2)_j} = - \frac{1}{N(N-1)}$ (see Methods), this simplifies to:
\begin{equation} \label{eq:quadratic_form_mf_step2}
\avg{\mathbf{v}_2^T L^{(2)} \mathbf{v}_2} \approx \avg{k^{(2)}_i} + \avg{a^{(2)}_{ij}} .
\end{equation}
Defining the expected connection probability of the $2$-path network as $p^{(2)} \equiv \avg{a^{(2)}_{ij}}$, and noting that the expected $2$-path degree is $\avg{k^{(2)}} = (N-1) p^{(2)}$, we arrive at:
\begin{equation} \label{eq:quadratic_form_mf_final}
\avg{\mathbf{v}_2^T L^{(2)} \mathbf{v}_2} \approx N p^{(2)}
\end{equation}
Substituting this mean-field result back into Eq.~\eqref{eq:zeta_perturbed}, the expected indirect influence due to $2$-paths is approximately:
\begin{equation} \label{eq:zeta_mf_2path}
\zeta \approx 2^{-\alpha} p^{(2)}.
\end{equation}
The expected $2$-path connection probability $p^{(2)}$ can be expressed in terms of the direct expected connection probability $p$ of the underlying network. A $2$-path exists between nodes $i$ and $j$ if they are not directly connected ($a_{ij} = 0$) but there exists at least one intermediate node $k$ such that $a_{ik} = 1$ and $a_{kj} = 1$. Using a mean-field argument for the existence probability over the $N-2$ potential intermediate nodes:
\begin{equation} \label{eq:p2_derivation}
\begin{split}
p^{(2)} = P \{ a^{(2)}_{ij} = 1 \} & = P \{ a_{ij} = 0 \} P\{ \exists k \in G \, | \, a_{ik} a_{kj} = 1 \} \\
& =  P \{ a_{ij} = 0 \} \left( 1 - P\{ \forall k \in G \, | \, a_{ik} a_{kj} = 0 \} \right) \\
& \approx  P \{ a_{ij} = 0 \} \left( 1- ( 1 - P\{ k \in G \, | \, a_{ik}a_{kj} = 1 \} )^{N-2} \right)\\
& = (1 - p) \left(1 - \left( 1 - p^2 \right)^{N-2}\right)
\end{split}
\end{equation}
This provides an explicit expression for $p^{(2)}$ based on the network parameters $N$ and $p$.

The perturbation and mean-field approach can be extended recursively to account for contributions from $d$-paths with $d > 2$. Each $d$-path term contributes approximately $d^{-\alpha} p^{(d)}$ to the indirect influence $\zeta$. The $d$-path expected connection probability $p^{(d)}$ can be approximated similarly to $p^{(2)}$, considering the condition that no shorter path exists between the nodes and that at least one path of length $d$ exists (see Methods): \begin{equation} \label{eq:pd_derivation}
p^{(d)} \approx \left( 1 - \sum_{i=1}^{d-1} p^{(i)} \right) \left( 1- \left( 1-p^d \right)^{\binom{N-2}{d-1}} \right).
\end{equation}

Therefore, the total indirect influence $\zeta$ can be approximated by summing the contributions from all relevant path lengths ($d = 2, \dots, D$):
\begin{equation} \label{eq:zeta_mf_general}
\zeta(N, p, \alpha) \approx \sum_{d=2}^{D} d^{-\alpha}  \left( 1 - \sum_{i=1}^{d-1} p^{(i)} \right) \left( 1- \left( 1-p^d \right)^{\binom{N-2}{d-1}} \right).
\end{equation}

%%%%%%%%%%%%%%%%%%%%%%%%%%%%%%%%%%%%%%%%%%%%%%%%%%%%%%%%%%%%%%%%%%%%
% FIGURE 2
\begin{figure}[tb!]
\centering
\begin{tabular}[t]{cc}
    ER networks & RR networks \\
    \includegraphics[width=0.4\textwidth]{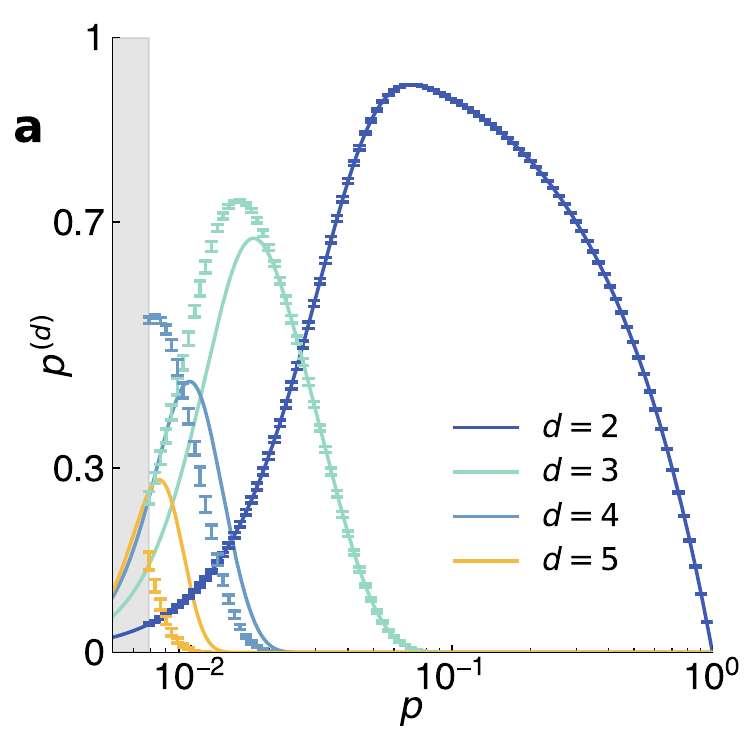}\label{fig2:a} &
    \includegraphics[width=0.4\textwidth]{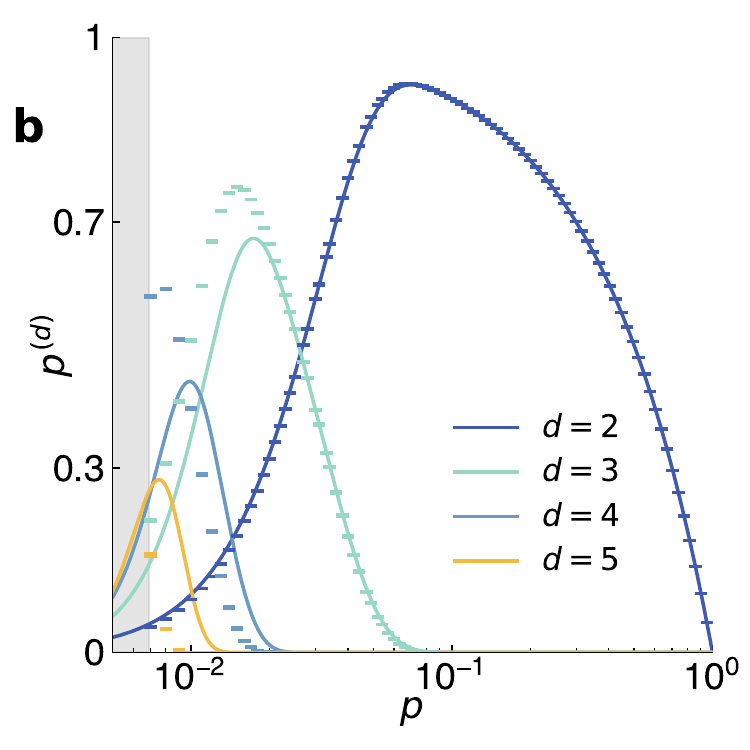}\label{fig2:b}
\end{tabular}
\caption{
\fontfamily{cmss}\selectfont{\textbf{Expected connection probabilities of $\bm{d}$-path networks relative to the 1-path network}. Expected connection probabilities for different $d$-path networks as a function of the expected 1-path connection probability, with ${N = 1000}$. \textbf{a}~Erd\H{o}s-R\'enyi (ER) networks and \textbf{b}~Random Regular (RR) networks. Each point corresponds to the average over 50 different networks with error bars of one standard deviation, and the solid lines correspond to their respective analytical expression. The gray shaded range indicates the values of ${p < \log N / N}$, corresponding to the values where the network is almost surely disconnected \cite{erdHos1961strength}, and therefore are not considered.
}}
\label{fig:fig2}
\end{figure}
%%%%%%%%%%%%%%%%%%%%%%%%%%%%%%%%%%%%%%%%%%%%%%%%%%%%%%%%%%%%%%%%%%%%

Figure~\ref{fig:fig2} shows the behavior of these calculated expected $d$-path probabilities $p^{(d)}$ as a function of the direct connection probability $p$ for both Erd\H{o}s-R\'enyi (ER) and Random Regular (RR) networks. The theoretical predictions are compared with results obtained from direct numerical simulations. The plots demonstrate that the recursive mean-field calculation accurately captures the dependence of $p^{(d)}$, especially for smaller values of $d$. As $d$ increases, the accuracy of the mean-field approximation slightly decreases, likely due to increasing correlations between paths that are ignored in the calculation.

Therefore, Eq.~\eqref{eq:zeta_mf_general}, combined with the recursive calculation of $p^{(d)}$ validated in Figure~\ref{fig:fig2}, provides a theoretical framework to predict the total indirect influence $\zeta$ based on the structural network properties and the influence decay parameter $\alpha$. \\

\noindent \textbf{Phase Diagram for Indirect Influence} \\
\noindent Having derived the theoretical approximation for the indirect influence, we now test its validity by comparing it against direct numerical simulations across a wide range of network parameters. We computed $\zeta$ numerically for ensembles of Erd\H{o}s-R\'enyi (ER) and Random Regular (RR) networks by calculating the second smallest eigenvalues $\hat{\lambda}_2$ and $\lambda_2$ of the full path Laplacian $\hat{L}$ and the standard Laplacian $L$, respectively.

%%%%%%%%%%%%%%%%%%%%%%%%%%%%%%%%%%%%%%%%%%%%%%%%%%%%%%%%%%%%%%%%%%%%
% FIGURE 3
\begin{figure}[tb!]
\centering
\begin{tabular}[t]{cc}
    ER networks & RR networks \\
    \includegraphics[width=0.45\textwidth]{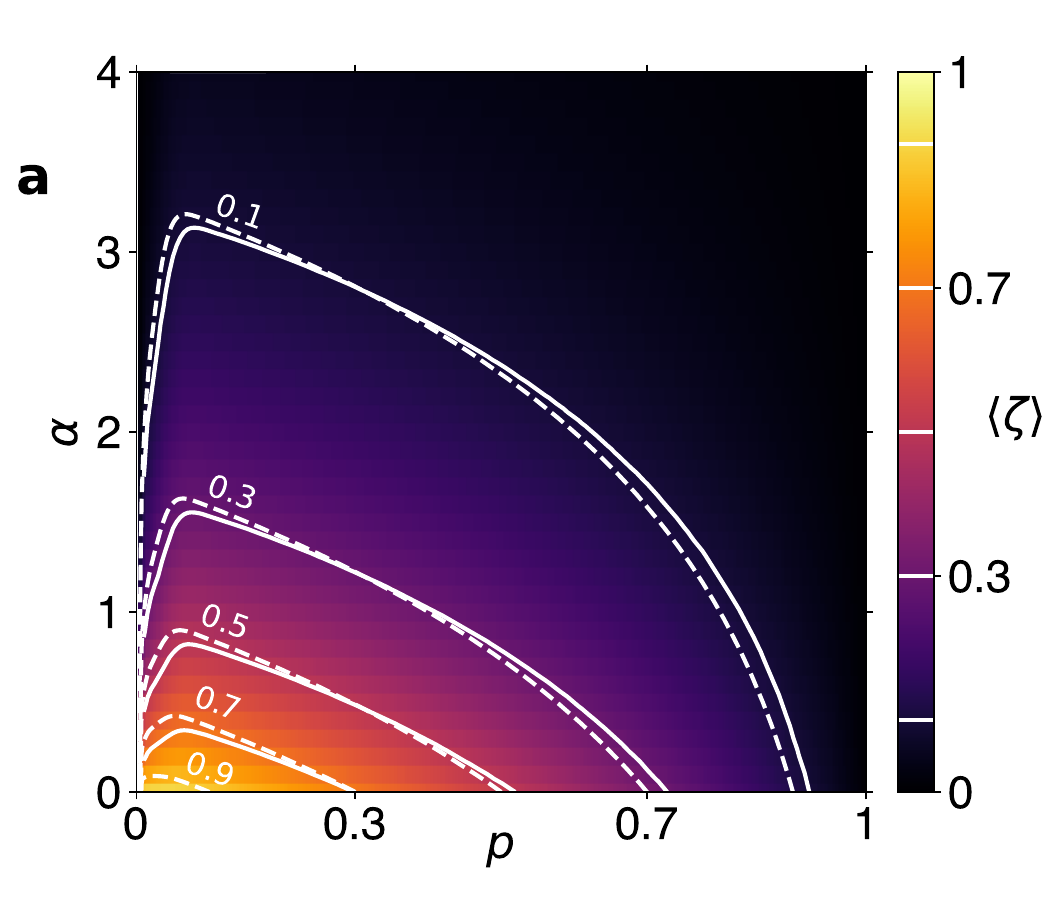}\label{fig3:a} &      \includegraphics[width=0.45\textwidth]{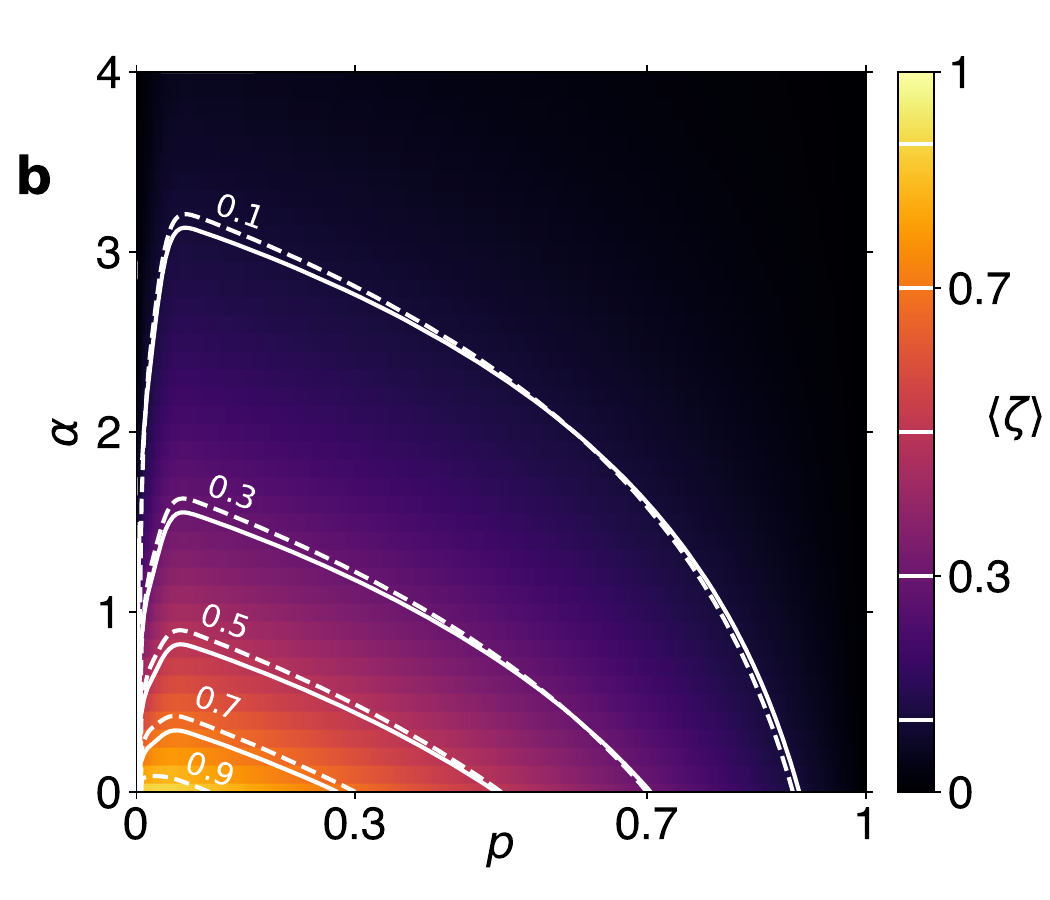}\label{fig3:b}
\end{tabular}
\caption{
\fontfamily{cmss}\selectfont{\textbf{Average indirect influence across parameter space}. Average indirect influence computed over all possible $d = 1, \dots, D$ paths for networks with $N = 1000$. \textbf{a}~Erd\H{o}s-R\'enyi (ER) networks. \textbf{b}~Random Regular (RR) networks. Solid white lines correspond to the contour levels of the color map that corresponds to the simulations, and the dashed white lines correspond to the theoretical contour levels. We compute 100 different networks for each combination of $\alpha$ ranging from 0 to 4 in steps of 0.1 and $p$ ranging from $0$ to $1$ in steps of 0.001.
}}
\label{fig:fig3}
\end{figure}
%%%%%%%%%%%%%%%%%%%%%%%%%%%%%%%%%%%%%%%%%%%%%%%%%%%%%%%%%%%%%%%%%%%%

Figure~\ref{fig:fig3} presents the phase diagram of the average indirect influence $\zeta$ in the parameter space defined by the connection probability $p$ and the decay exponent $\alpha$, for both ER and RR networks with $N=1000$. The agreement between the theoretical predictions and the simulation results is remarkably good across the entire phase space for both network types. This alignment holds even for small values of $\alpha$, which are technically outside the $\alpha \gg 1$ regime where the perturbation theory was initially applied. This indicates that the mean-field approximation for the expected $d$-path probabilities $p^{(d)}$ and the additive contribution assumption in Eq.~\eqref{eq:zeta_mf_general} effectively capture the collective impact of indirect pathways on diffusion dynamics.

The phase diagram provides valuable insights into when indirect influence becomes significant. As expected, $\zeta$ decreases as $\alpha$ increases, reflecting the diminished contribution of longer paths when influence decays rapidly. For a fixed $\alpha$, $\zeta$ generally decreases with increasing $p$, as denser networks offer less potential indirect pathways. This phase diagram serves as a practical tool to determine, based on network structure and the assumed decay mechanism, whether incorporating indirect connections is likely to substantially alter the system's diffusion timescale compared to a standard direct-only model. Furthermore, the framework provides a quantitative measure, $\zeta$, of the magnitude of this effect. \\

\noindent \textbf{Analysis of the Uniform Weighting Limit ($\bm{\alpha = 0}$)} \\
\noindent The theoretical framework based on perturbation theory assumes $\alpha \gg 1$. We now investigate the opposite limit, $\alpha = 0$, where influence does not decay with distance, i.e., $d^{-\alpha} = 1$ for all $d$. In this limit, the path Laplacian becomes $\hat{L} = \sum_{d=1}^D L^{(d)}$ and the perturbation approach is no longer valid, requiring an alternative analysis.

If we consider only direct and $2$-path connections, the path Laplacian is $\hat{L} = L^{(1)} + L^{(2)}$. This structure resembles the Laplacian used in studies of diffusion on duplex networks \cite{gomez2013diffusion, torres2024structural}. Analogously, we can approximate the combined $1$-path and $2$-path network as a single effective network. Since the edge sets of the $1$-path and $2$-path networks are disjoint, by definition of $A^{(2)}$, we approximate this effective network as an Erd\H{o}s-R\'enyi network with $N$ nodes and an effective connection probability $p_\text{eff} = p + p^{(2)}$, where $p$ and $p^{(2)}$ are given by Eq.~\eqref{eq:p2_derivation}. The algebraic connectivity $\hat{\lambda}_2$ for this truncated $\alpha = 0$ case can then be approximated by the known results for the algebraic connectivity of ER and RR networks \cite{jamakovic2008robustness, mckay1981expected}. The indirect influence can be written as:
\begin{equation} \label{eq:zeta_alpha0_approx}
\zeta_{\alpha=0}^{\text{ER/RR}} = \frac{\lambda^{\text{ER}}_{2}(N, p + p^{(2)}) - \lambda^{\text{ER/RR}}_2(N, p)}{N}.
\end{equation}
The approximation for large ($N \gg 1$) unweighted RR networks of degree~$k=p(N-1)$ is:
\begin{equation} \label{RR}
  \lambda^\text{RR}_2(k)
  \approx
  k - 2\sqrt{k-1}\,.
\end{equation}
And for large unweighted ER networks is:
\begin{equation} \label{ER}
  \begin{split}
  \lambda^\text{ER}_2(N,p)
  \approx
  p (N - 1) &
  - \sqrt{2p(1 - p)(N - 1) \log N}
  \\
  & + \sqrt{\frac{(N - 1)p(1 - p)}{2 \log N}} \log \sqrt{2 \pi \log \left( \frac{N^2}{2\pi} \right)}
  \\
  & - \sqrt{\frac{(N - 1)p(1 - p)}{2 \log N}}\,\gamma\,,
\end{split}
\end{equation}
where $p$ is the connection probability of the nodes and $\gamma$ is the Euler–Mascheroni constant.

%%%%%%%%%%%%%%%%%%%%%%%%%%%%%%%%%%%%%%%%%%%%%%%%%%%%%%%%%%%%%%%%%%%%
% FIGURE 4
\begin{figure}[tb!]
\centering
\begin{tabular}[t]{cc}
    ER networks & RR networks \\
    \includegraphics[width=0.35\textwidth]{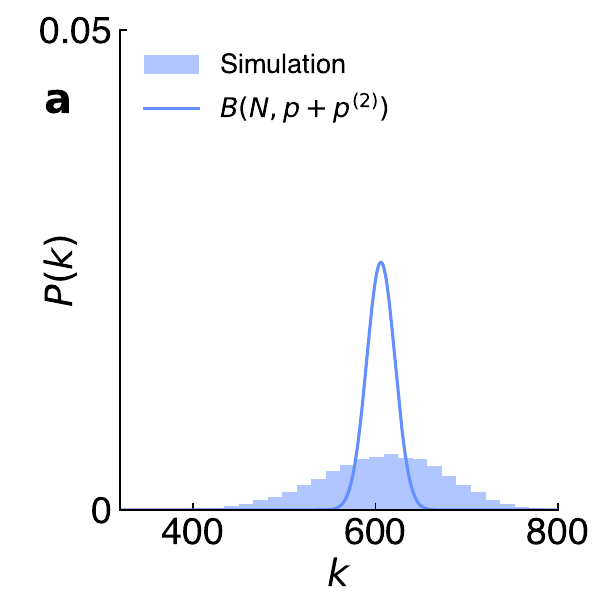} &      \includegraphics[width=0.35\textwidth]{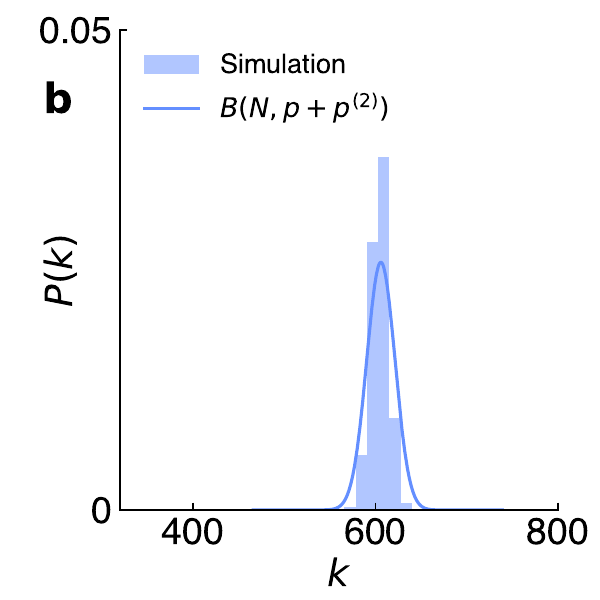}
\end{tabular}
\caption{
\fontfamily{cmss}\selectfont{\textbf{Degree distribution of the path network}. Degree distribution of the path network, $k = k^{(1)} + k^{(2)}$, which includes 1-path and 2-path links, averaged over 50 realizations of \textbf{a}~ER networks and \textbf{b}~RR networks. In both cases, $N = 1000$ and $\avg{k^{(1)}} = 30$. Solid lines represent the theoretical binomial distribution that an ER network should have, with $N$ nodes and a connection probability of $p + p^{(2)}$.
}}
\label{fig:fig4}
\end{figure}
%%%%%%%%%%%%%%%%%%%%%%%%%%%%%%%%%%%%%%%%%%%%%%%%%%%%%%%%%%%%%%%%%%%%

To assess the validity of this ER approximation for the combined 1-path and 2-path network, we examine its degree distribution in Figure~\ref{fig:fig4}. While the average degree matches the expectation, the shape of the distribution differs from the theoretical binomial distribution predicted for an equivalent ER graph. For ER base networks, Figure~\ref{fig:fig4}a, the combined network exhibits a broader tail, whereas for RR base networks, Figure~\ref{fig:fig4}b, the distribution is narrower. These differences indicate that the combined $1$-path and $2$-path network structure deviates from a pure ER graph, which may introduce inaccuracies in the $\hat{\lambda}_2$ approximation based on $\lambda^{\text{ER}}_2$.

If we consider the full path Laplacian at $\alpha = 0$, assuming the underlying network is connected, the path network effectively becomes the complete graph $K_N$. This is because for any two nodes $i$, $j$ in a connected graph, there exists a shortest path of some length $d \leq D$. The algebraic connectivity of the complete unweighted network is $\lambda_2 = N$ \ \cite{fiedler1973algebraic}. Therefore, in this limit, the indirect influence reaches its maximum possible value: \begin{equation} \label{eq:zeta_alpha0_full}
\zeta_{\alpha=0}^{\text{ER/RR}} = \frac{N - \lambda^{\text{ER/RR}}_2(N, p)}{N}.
\end{equation}

%%%%%%%%%%%%%%%%%%%%%%%%%%%%%%%%%%%%%%%%%%%%%%%%%%%%%%%%%%%%%%%%%%%%
% FIGURE 5
\begin{figure}[tb!]
\centering
\begin{tabular}[t]{cc}
    \fontfamily{cmss} ER networks & RR networks \\
    \includegraphics[width=0.45\textwidth]{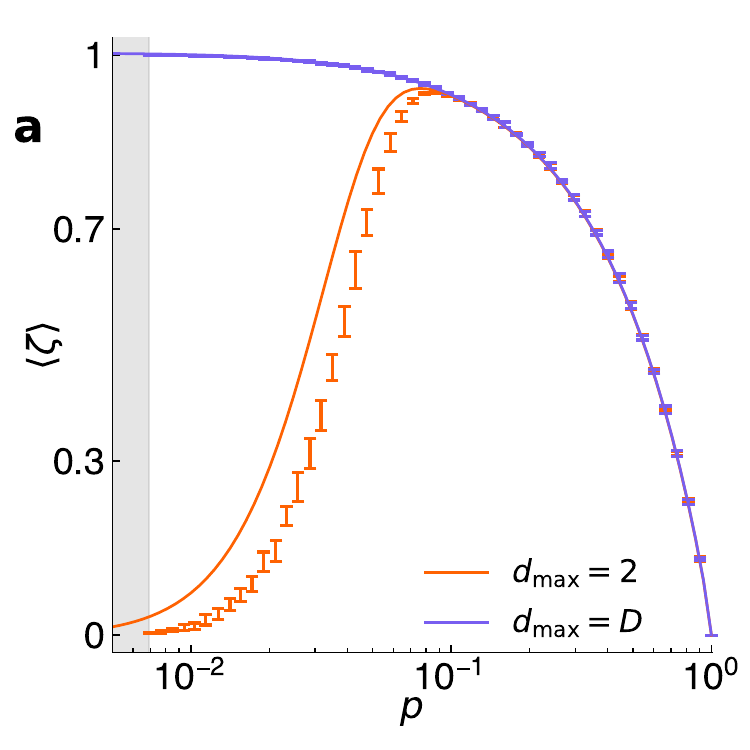}\label{fig5:a} &      \includegraphics[width=0.45\textwidth]{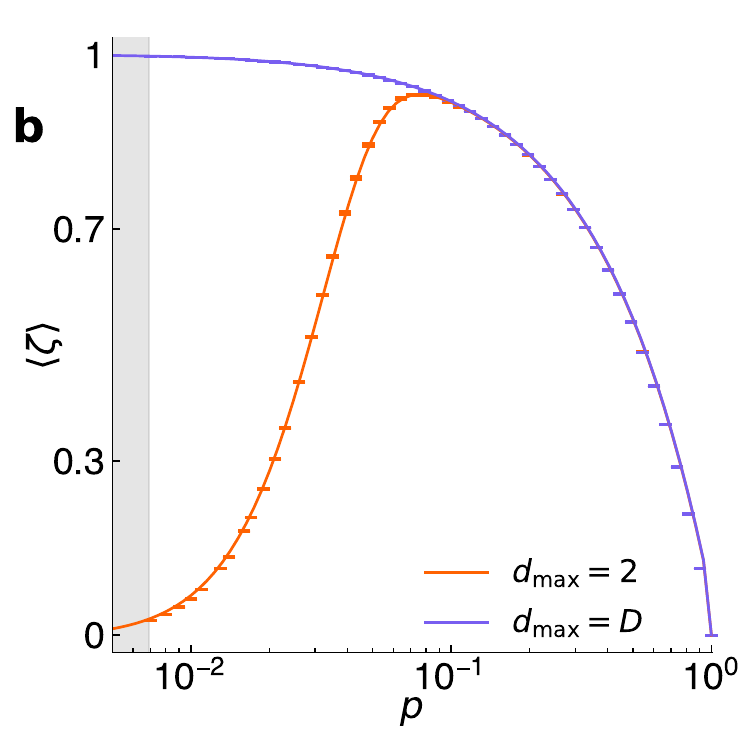}\label{fig5:b}
\end{tabular}
\caption{
\fontfamily{cmss}\selectfont{\textbf{Average indirect influence of the uniform weight limit}. Average indirect influence for \textbf{a}~Erd\H{o}s-R\'enyi networks and \textbf{b}~Random Regular networks as a function of the expected connection probability for the path Laplacian truncated at $d_{\max} = 2$ and $d_{\max} = D$. Each point corresponds to the average over 50 different networks with error bars of one standard deviation, and the solid lines correspond to the theoretical curves for the respective truncation. The gray shaded range indicates the values of $p < \log N / N$, corresponding to the values where the network is almost surely disconnected, and therefore are not considered.}}
\label{fig:fig5}
\end{figure}
%%%%%%%%%%%%%%%%%%%%%%%%%%%%%%%%%%%%%%%%%%%%%%%%%%%%%%%%%%%%%%%%%%%%

Figure~\ref{fig:fig5} compares the theoretical approximations for $\alpha = 0$ with simulation results. Overall, the results are in strong agreement. However, for ER networks, there are slight deviations that align with the differences in the degree distributions shown in Figure~\ref{fig:fig4}. This analysis highlights that while the $\alpha = 0$ case represents maximal indirect influence, accurately predicting it requires either considering the full path network or carefully accounting for the structural specifics of truncated path networks beyond simple ER approximations. \\

\noindent \textbf{Structural Phase Transition of Indirect Connections} \\
\noindent The preceding analyses show that the significance of indirect influence $\zeta$ is closely related to the prevalence of $d$-paths, particularly those corresponding to non-negligible weights $d^{-\alpha}$. The sharp changes observed in the $p^{(d)}$ curves and the behavior of $\zeta$ at low $p$ suggest the existence of an underlying structural transition related to the emergence of these indirect connections. We now characterize this phenomenon more formally.

Inspired by the concept of connectivity transitions in random graphs \cite{erdHos1961strength}, we focus on the completeness of the network formed by considering both direct and indirect paths. For simplicity, let's first consider the network formed by only $1$-paths and $2$-paths, ignoring the weighting $d^{-\alpha}$ and focusing solely on the presence or absence of connections. The density of connections in this combined network can be quantified by its completeness $\nu$, defined as the probability that a randomly chosen pair of nodes $i$ and $j$ is connected by either a 1-path or a 2-path:
\begin{equation} \label{eq:completeness_d2}
    \nu  = p + p^{(2)}
\end{equation}
Substituting the expression of $p^{(2)}$ from Eq.~\ref{eq:p2_derivation}, we get:
\begin{equation} \label{eq:completeness_d2_expanded}
\nu = p + (1 - p) \left( 1- \left( 1 - p^2 \right)^{N-2} \right).
\end{equation}
We consider this completeness $\nu$ as an order parameter, with the direct expected connection probability $p$ acting as the control parameter. We look for a phase transition, marked by a critical value $p_c(N)$, where the probability of the combined network being fully connected sharply transitions from $0$ to $1$ as $N \rightarrow \infty$, i.e.,
\begin{equation*}
    \begin{split}
        p < p_c &\Rightarrow P \{ \nu = 1 | N \rightarrow \infty \} = 0, \\
        p > p_c &\Rightarrow P \{ \nu = 1 | N \rightarrow \infty \} = 1.
    \end{split}
\end{equation*}

The probability that the combined $1$-path and $2$-path network is a complete network is equivalent to the probability that no pair $i$, $j$ lacks both a $1$-path and $2$-path connection. Assuming independence between pairs for large $N$:
\begin{equation}
\label{eq:complet}
    \begin{split}
P\{\nu = 1\} & = 1 - P\{ \exists i,j \, | \, a_{ij} = a^{(2)}_{ij} = 0 \} \\
& = 1 - \left(1 - \left( p + p^{(2)} \right) \right) = p + p^{(2)} \\
& = p + (1-p)\left( 1- \left( 1 - p^2 \right)^{N-2} \right) \\
& = 1 - \left( 1 - p^2 \right)^{N} \\
& = 1 - e^{-p^2 N}
    \end{split}
\end{equation}
In the last step, we use the property that $\lim_{N \to \infty} \left(1 - \frac{a}{N}\right)^N = e^{-a}$. Therefore, from this development it is immediate to see that $p_c(N)=N^{-1/2}$ is a threshold function. This threshold marks a structural phase transition. Below $p_c(N)$, the network is sparse in terms of 2-paths, while above it, the network rapidly becomes dense with $2$-paths, causing the completeness $\nu$ to approach $1$ and significantly boosting the indirect influence.

%%%%%%%%%%%%%%%%%%%%%%%%%%%%%%%%%%%%%%%%%%%%%%%%%%%%%%%%%%%%%%%%%%%%
% FIGURE 6
\begin{figure}[tb!]
\centering
\begin{tabular}[t]{cc}
    \includegraphics[width=0.44\textwidth]{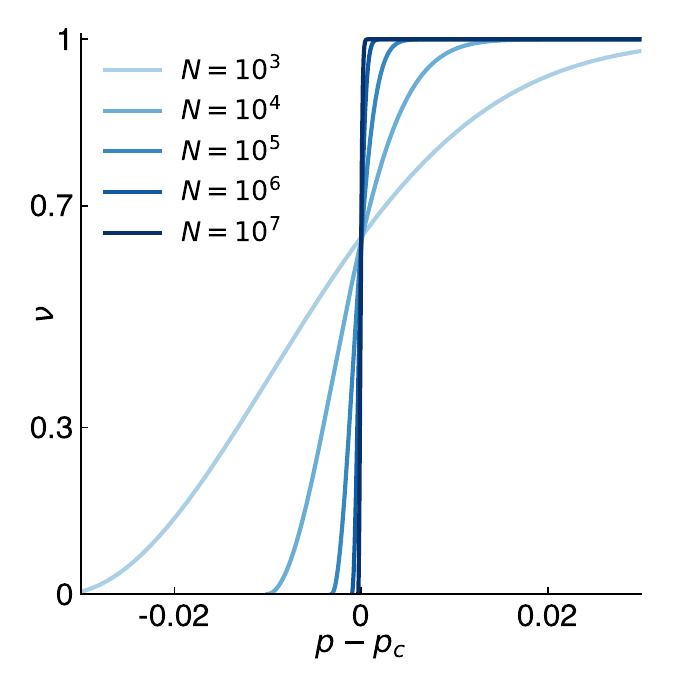}
\end{tabular}
\caption{
\fontfamily{cmss}\selectfont{\textbf{Completeness of the path network}. Completeness of the path network, accounting for 1-path and 2-path links, as a function of the expected connection probability, rescaled with the respective critical threshold for different values of $N$.
}}
\label{fig:fig6}
\end{figure}
%%%%%%%%%%%%%%%%%%%%%%%%%%%%%%%%%%%%%%%%%%%%%%%%%%%%%%%%%%%%%%%%%%%%

Figure~\ref{fig:fig6} visually confirms this transition by plotting the completeness against the expected connection probability rescaled by the critical threshold.

This analysis can be generalized to consider the emergence of $d$-paths for $d > 2$. The probability that a $d$-path exists between nodes $i$, $j$ given that no shorter path exists, is controlled by terms approximating $(1-p^d)^{N^{d-1}}$, therefore $P\{\nu = 1\} \propto e^{-p^d N^{d-1}}$. Since each successive exponential term decays faster, the general threshold function is
\begin{equation} \label{eq:pc_d_general}
p_{c}(N) = N^{-\frac{d-1}{d}}.
\end{equation}
This threshold function provides the critical value of the expected connection probability for the onset of this new structural phase transition involving up to $d$-path indirect connections. Notably, as $N$ increases, and with it the network diameter $D$, the maximum possible $d$ term in the path Laplacian also grows. For sufficiently large $d$, this critical value can fall below the connectivity threshold $\log(N)/N$, causing this structural phase transition to disappear.

\section*{\large{Discussion}}

This work provides a quantitative framework for analyzing the impact of indirect pathways on network diffusion. The central results, including the indirect influence parameter $\zeta$, its predictive phase diagram, and the critical threshold for the onset of $d$-paths, $p_c(N)$, offer a practical method for assessing the significance of non-local interactions. The theory, derived for analytically tractable random networks, is validated by numerical simulations, confirming that indirect pathways emerge as a collective structural phenomenon rather than a gradual effect.

The theoretical approach relies on mean-field approximations that are well-suited for random graphs. This scope invites future work on networks with more complex topologies, such as those with high clustering or community structure, where structural correlations would likely alter the quantitative predictions. A key implication of our framework is the potential for empirical measurement of $\zeta$. Observing the global diffusion timescale of a process on a network with a known structure would allow for the calculation of an empirical $\zeta$. This value could then be used to test the model and estimate system-specific parameters, such as the influence decay rate $\alpha$, bridging the gap between theory and real-world data.

The model employs a power-law decay of influence $d^{-\alpha}$, a flexible choice corresponding to a Mellin transform. While other functional forms, like exponential decay, are possible, the qualitative features of the phase transition are expected to be robust. In conclusion, this study's findings provide a rigorous guide for model selection in network science, with direct applications in fields such as epidemiology and the study of social influence.

\section*{\large{Methods}}

\noindent \textbf{Indirect Diffusion Model} \\
\noindent We model diffusion dynamics on a network represented as an unweighted, connected graph $G = (V, E)$, where $V$ is the set of $N$ nodes (indexed $i = 1, ..., N$) and $E$ is the set of edges representing direct connections between nodes. To incorporate the influence of interactions beyond direct neighbors, we employ the $d$-path Laplacian framework \cite{estrada2012path}.

This framework makes use the $d$-path adjacency matrices, $A^{(d)} \in \{0,1\}^{N \times N}$, that account for all shortest paths of length $d = 1,\ldots, D$ connecting nodes in the network, where $D$ is the diameter of the network. They are defined as
\begin{equation}
        (A^{(d)})_{ij} = a^{(d)}_{ij} = \begin{cases}
        1 & \text{if } \dist(i,j) = d \\ 0 & \text{otherwise}
    \end{cases},
\end{equation}
where $\dist(i,j)$ denotes the shortest path distance between nodes $i$ and $j$. Note that $A^{(1)}$ corresponds to the standard adjacency matrix $A$.

By extending the direct diffusion formalism \cite{masuda2017random}, the diffusion equations for a network with indirect connections are
\begin{equation} \label{eq:idiff}
    \frac{d x_i}{dt} = \gamma^{(1)} \sum_{j=1}^N a^{(1)}_{ij} (x_j - x_i) + \cdots + \gamma^{(D)} \sum_{j=1}^N a^{(D)}_{ij} (x_j - x_i) ,
\end{equation}
where $x_i$ is the state of node $i$, and $\gamma^{(d)} > 0$ is the diffusion rate along $d$-path connections. It is natural to assume that the diffusion rate decays as a function of the distance $d$ between nodes. This choice can be conveniently tuned depending on the problem at hand. Moreover, it is convenient to set the time scale of the process with $\gamma^{(1)} = 1$. Without loss of generality, we consider an exponential decay of the form $\gamma^{(d)} = d^{-\alpha}$ where $\alpha$ is the new general control parameter. This formulation corresponds to a Mellin transformation \cite{estrada2012path}.

The $d$-path Laplacian $L^{(d)} \in \mathds{R}^{N \times N}$ is defined as
\begin{equation}
    (L^{(d)})_{ij} = \begin{cases}
    -1 & \text{if } \dist(i,j) = d \\ k^{(d)}_i & \text{if } i = j \\ 0 & \text{otherwise}
    \end{cases}
\end{equation}
where ${k^{(d)}_i = (\mathbf{1}^T A^{(d)})_i}$ is the degree of node $i$ in the $d$-path adjacency matrix. This definition corresponds to the matrix form
\begin{equation}
    L^{(d)} = K^{(d)} - A^{(d)} ,
\end{equation}
where $K^{(d)}$ is the diagonal matrix of node degrees for $A^{(d)}$.
To capture diffusion across paths of all lengths, we define the path Laplacian as
\begin{equation} \label{eq:pathlaplacian}
    \hat{L} = \sum_{d=1}^{D} d^{-\alpha} L^{(d)} .
\end{equation}
Using the path Laplacian definition, Eq.~\eqref{eq:idiff} can be compactly written as
\begin{equation}
    \Dot{\mathbf{x}} = - \hat{L} \mathbf{x} .
\end{equation}

\noindent \textbf{Perturbation Theory Derivation} \\
\noindent To derive the first-order perturbation term of the $d$-path Laplacian, we recall the perturbation framework
\begin{equation}
    \hat{L} \hat{\mathbf{v}}_2 = \hat{\lambda}_2 \hat{\mathbf{v}}_2 \rightarrow
    \begin{cases}
    \hat{\lambda}_2 = \lambda_2 + \epsilon \lambda' + o(\epsilon^2) \\
    \hat{\mathbf{v}}_2 = \mathbf{v}_2 + \epsilon \mathbf{v}' + o(\epsilon^2)
    \end{cases} .
\end{equation}
By introducing the perturbations into the eigenvalue equation, we obtain:
\begin{equation}
    \left(L + \epsilon L^{(2)}\right)(\mathbf{v}_2 + \epsilon \mathbf{v}') = (\lambda_2 + \epsilon \lambda')(\mathbf{v}_2 + \epsilon \mathbf{v}').
\end{equation}
After expanding the equality and ignoring second-order terms, we get
\begin{equation}
    L \mathbf{v}' + L^{(2)} \mathbf{v}_2 = \lambda_2 \mathbf{v}' + \lambda' \mathbf{v}_2.
\end{equation}
We then multiply both sides of the equation by $\mathbf{v}^T_2$ and simplify
\begin{equation}
    \mathbf{v}_2^T L^{(2)} \mathbf{v}_2 = \mathbf{v}_2^T \lambda' \mathbf{v}_2.
\end{equation}
Lastly, since the eigenvectors are orthonormal, we end up with
\begin{equation}
    \lambda' = \mathbf{v}_2^T L^{(2)} \mathbf{v}_2.
\end{equation}

\noindent \textbf{Mean-Field Approximation Details} \\
\noindent To derive an analytical expression for the indirect influence, we used the first and second moments of the entries of the eigenvector $\mathbf{v}_2$. These moments are obtained from the properties $\sum_{i=1}^N (\mathbf{v}_2)_i^2 = 1$ and $\sum_{i=1}^N (\mathbf{v}_2)_i = 0$, which correspond to the normalization  of $\mathbf{v}_2$, and orthogonality of $\mathbf{v}_2$ and $\mathbf{v}_1=\mathbf{1}$, respectively. To compute the first moment, we begin by taking the expectation of both sides of the first property, yielding:
\begin{equation}
    \sum_{i=1}^N \avg{(\mathbf{v}_2)_i^2} = 1.
\end{equation}
If we assume that the eigenvector entries are identically independent, we get
\begin{equation}
    \avg{(\mathbf{v}_2)_i^2} = \frac{1}{N}.
\end{equation}
For the second moment, we start from
\begin{equation}
      \sum_{i=1}^N (\mathbf{v}_2)_i = 0\ \Rightarrow \ 0 = \left( \sum_{i=1}^N (\mathbf{v}_2)_i \right) \left( \sum_{j=1}^N (\mathbf{v}_2)_j \right).
\end{equation}
Then, by isolating the diagonal terms, we obtain:
\begin{equation}
\begin{split}
    0 = \sum_{i=1}^N \sum_{j=1}^N (\mathbf{v}_2)_i (\mathbf{v}_2)_j & = \sum_{i=1}^N (\mathbf{v}_2)_i^2 + \sum_{i=1}^N \sum_{j \neq i}^N (\mathbf{v}_2)_i (\mathbf{v}_2)_j \\
    & =  1 + \sum_{i=1}^N \sum_{j \neq i}^N (\mathbf{v}_2)_i (\mathbf{v}_2)_j.
\end{split}
\end{equation}
We used the first eigenvector property to evaluate the first term after splitting the summations. Finally, after taking the expected value on both sides, we get:
\begin{equation}
    \avg{ (\mathbf{v}_2)_i (\mathbf{v}_2)_j } = -\frac{1}{N(N-1)}.
\end{equation}

\noindent \textbf{$\bm{d}$-path Expected Connection Probability} \\
\noindent We derive the general $d$-path expected connection probability in terms of the expected connection probability of the base network. The existence of a $d$-path connection requires two conditions: the absence of all shorter $d'$-path connections for $d' < d$, and the existence of at least one $d$-path between the nodes. Therefore:
\begin{widetext}
\begin{equation}
        p^{(d)} = P \{ a^{(d)}_{ij} = 1 \} = P\{ a_{ij} = 0,... ,a^{(d-1)}_{ij} = 0 \} P\{ \exists k_{1},...,k_{d} \in G \, | \, a_{ik_{1}}...a_{k_{d}j} = 1 \}.
\end{equation}
\end{widetext}
The second term is equivalent to one minus the probability of the absence of all possible $d$-path connections. Using the mean-field approximation, which assumes independence between the absence of these paths, we obtain:
\begin{widetext}
\begin{equation}
\begin{split}
    p^{(d)} & = P\{ a_{ij} = 0,... ,a^{(d-1)}_{ij} = 0 \} ( 1- P\{ \forall k_{1},...,k_{d-1} \in G \, | \, a_{ik_{1}}...a_{k_{d-1}j} = 0 \} ) \\
    & \approx P\{ a_{ij} = 0,... ,a^{(d-1)}_{ij} = 0 \} \left( 1- ( 1 - P\{ k_{1},...,k_{d-1} \in G \, | \, a_{ik_{1}} ... a_{k_{d-1}j} = 1 \} ) ^{\binom{N-2}{d-1}} \right).
\end{split}
\end{equation}
\end{widetext}

Finally, we can express the first probability term in terms of the previously derived expected connection probabilities, while the second term simplifies to $p^d$, since the existence of a $d$-path connection requires the presence of each direct link along the path.
\begin{equation}
    p^{(d)} \approx \left(1 - \sum_{i=1}^{d-1} p^{(i)}\right)
    \left(1- \left( 1-p^d \right)^{\binom{N-2}{d-1}} \right).
\end{equation}
Since the $p^{(i)}$ probabilities can be recursively expressed in terms of $p$, each $p^{(d)}$ ultimately depends only on the expected connection probability of the direct links. \\

\noindent \textbf{Numerical Simulations} \\
\noindent Numerical results for comparison with analytical predictions were obtained by generating ensembles of Erd\H{o}s-R\'enyi (ER) and Random Regular (RR) graphs using standard algorithms. For each set of parameters ($N$, $p$), results were typically averaged over multiple independent network realizations, as indicated in the figure captions. The shortest path distances required for constructing $d$-path matrices were computed using the Bellman–Ford algorithm, and the eigenvalues of both the standard Laplacian $L$ and the path Laplacian $\hat{L}$ were calculated using the Arpack numerical solver.

\vspace{4.ex}
\begingroup
\fontsize{8pt}{9pt}\selectfont

\section*{\large{Data availability}}
Data sharing not applicable to this article as no datasets were generated or analyzed during the current study.

\section*{\large{Code availability}}
The code for replicating the results and figures is availabale at \href{https://github.com/llui2/indirect-diffusion}{github.com/llui2/indirect-diffusion}.

\endgroup

%\bibliographystyle{unsrt}
%\bibliography{refs}

\begingroup
\fontsize{8pt}{9pt}\selectfont

\section*{\large{Acknowledgments}}
L.T.-H.\ acknowledges financial support from Diputació de Tarragona and Universitat Rovira i Virgili, Spain (2023PMF-PIPF-21). This work was supported by Ministerio de Ciencia e Innovación, Spain (PID2021-128005NB-C21, RED2022-134890-T), Generalitat de Catalunya, Spain (2021SGR-633), MICIU/AEI/10.13039/501100011033 FEDER EU (PID2022-142600NB-I00), and Universitat Rovira i Virgili, Spain (2023PFR-URV-00633).

\section*{\large{Author contributions}}
L.T.-H.\ contributed to the conceptualization, formal analysis, investigation, methodology, software development, writing the original draft, and manuscript review and editing.
J.D.\ contributed to the conceptualization, funding acquisition, and manuscript review and editing.
S.G.\ contributed to the conceptualization, formal analysis, funding acquisition, and manuscript review and editing.

\section*{\large{Competing interests}}
The authors declare no competing interests.

\endgroup

\end{document}